\documentclass[%
 reprint,
 amsmath,amssymb,
 aps,
]{revtex4-2}
\usepackage{graphicx}% Include figure files
\usepackage{dcolumn}% Align table columns on decimal point
\usepackage{bm}% bold math
\usepackage[utf8]{inputenc}
\usepackage[T1]{fontenc}
\usepackage{xcolor}
\usepackage{gensymb}
\usepackage[colorlinks=true,allcolors=blue]{hyperref}

\newcommand{\Rc}{\boldsymbol{\mathcal{R}}}
\newcommand{\Phit}{\boldsymbol{\Phi}}
\newcommand{\rhos}{\rho_{\Rc,\Phit}}

\begin{document}

\preprint{APS/123-QED}

%\title{Phonon mediated superconductivity in the symmetry-broken ground state of Kagome CsV$_3$Sb$_5$}% Force line breaks with \\
\title{Symmetry-broken ground state and phonon mediated superconductivity in Kagome CsV$_3$Sb$_5$}% Force line breaks with \\

\author{Manex Alkorta$^\mathrm{1,2}$}
\author{Martin Gutierrez-Amigo$^\mathrm{3}$}
\author{\DJ or\dj e Dangi\'c$^\mathrm{1,2}$}
\author{Chunyu Guo$^\mathrm{4}$}
\author{Philip J. W. Moll$^\mathrm{4}$}
\author{Maia G. Vergniory$^\mathrm{5,6,7}$}
\author{Ion Errea$^\mathrm{1,2,5}$}
 \affiliation{$^1$Centro de Física de Materiales (CFM-MPC), CSIC-UPV/EHU, Donostia, 20018, Spain}
 \affiliation{$^2$Department of Applied Physics, University of the Basque Country (UPV/EHU), Donostia, 20018, Spain}
 %\altaffiliation[Also at ]{Physics Department, XYZ University.}%Lines break automatically or can be forced with \\}%
\affiliation{$^3$
Department of Applied Physics, Aalto University School of Science, FI-00076 Aalto, Finland}%
\affiliation{$^4$Max Planck Institute for the Structure and Dynamics of Matter, Hamburg, 22761, Germany}
 \affiliation{$^5$Donostia International Physics Center (DIPC), Donostia, 20018, Spain}
\affiliation{$^6$Département de Physique et Institut Quantique, Université de Sherbrooke, Sherbrooke, J1K 2R1 Québec, Canada.}
\affiliation{$^7$Regroupement Qu\'eb\'ecois sur les Mat\'eriaux de Pointe (RQMP), Quebec H3T 3J7, Canada}

\date{\today}% It is always \today, today,

\begin{abstract}
The newly discovered family of non-magnetic Kagome metals AV$_3$Sb$_5$ (A=K,Rb,Cs) provides a unique platform for exploring the interplay between charge density wave (CDW) order, superconductivity, non-trivial topology, and spontaneous time-reversal symmetry breaking. 
Although characterizing the CDW phase is essential for understanding and modeling these exotic phenomena, its nature remains unresolved.
In this work, we employ first-principles free-energy calculations, accounting for both ionic kinetic energy and anharmonic effects, to resolve the atomistic phase diagram of CsV$_3$Sb$_5$ and its charge ordering structure. 
Our results uncover that the CDW ground state is formed by reconstructed vanadium Kagome layers in a triangular-hexagonal pattern, featuring energetically degenerate different stacking orders. This accounts for the various out-of-plane modulations observed experimentally and supports the coexistence of multiple domains.
The discovered symmetry-broken ground state is consistent with the absence of any electronic anisotropy in transport experiments. 
By combining anharmonic phonons with the calculation of electron-phonon matrix elements, we predict a superconducting critical temperature for the CDW phase in agreement with experiments, showing that superconductivity is phonon mediated.  
These findings not only resolve a long-standing structural puzzle, but also clarify the impact of the CDW in superconductivity, highlighting its fundamental importance in shaping the low-temperature quantum phase diagram of Kagome metals.
\end{abstract}

\maketitle

The Kagome lattice has been a subject of interest across various fields of physics since its introduction in 1951~\cite{syoziStatisticsKagomeLattice1951}. 
Composed of corner-sharing triangles arranged in a hexagonal cell, it exhibits rich electronic behavior, including frustrated magnetism~\cite{sachdevKagomeTriangularlatticeHeisenberg1992, kangTopologicalFlatBands2020}, flat bands~\cite{kangDiracFermionsFlat2020, bilitewskiDisorderedFlatBands2018, calugaruGeneralConstructionTopological2022, kangTopologicalFlatBands2020}, and symmetry-protected Dirac cones~\cite{zhaoCascadeCorrelatedElectron2021, guoCorrelatedOrderTipping2024, mazinTheoreticalPredictionStrongly2014, ghimireTopologyCorrelationsKagome2020}, all of which give rise to topologically non-trivial phenomena~\cite{yinTopologicalKagomeMagnets2022, ghimireTopologyCorrelationsKagome2020}.
Among Kagome materials, the recently discovered non-magnetic AV$_3$Sb$_5$ family (A=K,Rb,Cs)~\cite{ortizNewKagomePrototype2019} stands out for hosting a remarkable combination of exotic quantum properties, such as electronic topology~\cite{huTopologicalSurfaceStates2022}, superconductivity~\cite{OrtizCsV3Sb5Superconductivity}, and spontaneous time-reversal symmetry breaking~\cite{jiangUnconventionalChiralCharge2021}, all within a still-unresolved charge-density wave (CDW) phase.
The lack of consensus regarding the nature of the CDW phase impedes a comprehensive understanding of these intertwined phenomena. 
Theoretical efforts to resolve the origin of superconductivity remain inconclusive, with both conventional (i.e., phonon-mediated)~\cite{wang_phonon-mediated_2023} and unconventional~\cite{wuNatureUnconventionalPairing2021, TanChargeDensityWaves2021, PhysRevB.111.134507} mechanisms being proposed. 
For instance, experimental evidence suggests that the coexistence of topological non-triviality and time-reversal symmetry breaking with superconductivity can lead to unusual effects, such as the superconducting diode effect and the emergence of Majorana modes~\cite{liang_three-dimensional_2021, le_superconducting_2024}.
Additionally, the origin of the observed chiral transport~\cite{guoSwitchableChiralTransport2022, guoDistinctSwitchingChiral2024} and the mechanisms underlying time-reversal symmetry breaking~\cite{fengChiralFluxPhase2021, yuEvidenceHiddenFlux2021} and superconductivity remain open questions.
Resolving the microscopic nature of the CDW and elucidating its role in the emergence of superconductivity thus represent key challenges at the forefront of current research.

All members of the AV$_3$Sb$_5$ family undergo a CDW phase transition at temperatures around 90 K. 
The nature of the emerging charge-ordered phases remains debated, particularly regarding its modulation pattern and symmetry preservation. While experimental evidences suggest a $2\times2\times2$ modulation for potassium and rubidium compounds~\cite{kautzschStructuralEvolutionKagome2023}, contradictory claims exist about the out of plane stacking of CsV$_3$Sb$_5$. Although most studies support a $2\times2\times2$ CDW reconstruction~\cite{zhaoCascadeCorrelatedElectron2021, kangChargeOrderLandscape2023, ratcliffCoherentPhononSpectroscopy2021, ningDynamicalDecodingCompetition2024, liObservationUnconventionalCharge2021, subiresOrderdisorderChargeDensity2023}, reports of a $2\times2\times4$ modulation~\cite{broylesEffectInterlayerOrdering2022, ortizFermiSurfaceMapping2021}, mixed domains of both~\cite{jinPhaseInterlayerShift2024, xiaoCoexistenceMultipleStacking2023, kautzschStructuralEvolutionKagome2023, stahlTemperaturedrivenReorganizationElectronic2022}, or even transitions between them~\cite{jinPhaseInterlayerShift2024} exist. The symmetry of the charge-reconstructed phase also remains uncertain. In K- and Rb-based structures, a six-fold symmetry breaking $\pi$-shifted triangular-hexagonal (TrH) ordering is well established~\cite{kautzschStructuralEvolutionKagome2023}. However, for CsV$_3$Sb$_5$, alternative stacking orders, such as mixed TrH and star-of-david (SoD)-like arrangements, seem to be necessary to interpret angle-resolved photoemission spectroscopy (ARPES)~\cite{kangChargeOrderLandscape2023} and X-ray measurements~\cite{kautzschStructuralEvolutionKagome2023}. 
In fact, the very existence of an intrinsic six-fold symmetry remains unclear. While scanning tunneling microscopy (STM)~\cite{zhaoCascadeCorrelatedElectron2021, nieChargedensitywavedrivenElectronicNematicity2022} observes a six-fold rotational symmetry breaking at the surface, corroborated by 120\textdegree-rotated nematic domains in magneto-optical Kerr measurements~\cite{xuThreestateNematicityMagnetooptical2022}, a recent electronic transport study on strain-free devices indicates that in-plane anisotropy is highly strain-sensitive, and vanishes in the absence of external perturbations~\cite{guoCorrelatedOrderTipping2024}. This raises the question of whether the six-fold symmetry breaking is an intrinsic property of the CDW itself.

The lack of consensus on the nature of the charge-reconstructed phase has motivated theoretical works in order to explain the origin of the CDW transition. The lack of evidence of phonon softening~\cite{chenAbsencePhononSoftening2024, subiresOrderdisorderChargeDensity2023, liuObservationAnomalousAmplitude2022, liObservationUnconventionalCharge2021}, along with the observed discontinuity in the lattice parameters at $T_{CDW}=94$ K~\cite{frachetColossalAxisResponse2024}, challenges the conventional soft-phonon driven transition so well-established in other CDW materials like transition-metal dichalcogenides (TMDs)~\cite{weberExtendedPhononCollapse2011, diegoVanWaalsDriven2021, biancoWeakDimensionalityDependence2020} and supports a first-order phase transition scenario. 
However, \emph{ab initio} harmonic phonons in the high-symmetry phase reveal dynamical instabilities that coincide with the experimentally observed in plane-doubling of the unit cell~\cite{subediHexagonaltobasecenteredorthorhombic4QCharge2022, siChargeDensityWave2022}, supporting the presence of soft-phonon physics. This also underlines how standard calculations, which  restrict to electronic Born-Oppenheimer (BO) energies and its second-order derivatives the analysis of the thermodynamic and dynamic stability of different CDW phases, completely break down. 
A recent study has indeed demonstrated that fully considering the ionic kinetic energy and anharmonicity is necessary to understand why the high-symmetry phase is stable and becomes the thermodynamic ground state above $T_{CDW}$~\cite{gutierrez-amigoPhononCollapseAnharmonic2024}. 

In this work, making use of first principle calculations including anharmonicity and ionic fluctuations within the stochastic self-consistent harmonic approximation (SSCHA)~\cite{monacelliStochasticSelfconsistentHarmonic2021,ErreaAnharmonicFreeEnergies2014,BiancoSecondOrderStructural2017,MonacelliPressureStressTensor2018}, we resolve the temperature dependent phase diagram of CsV$_3$Sb$_5$, unveiling its charge-ordered structure below $T_{CDW}$. We demonstrate that the TrH ordering of the Kagome vanadium plane is the only candidate for the low-symmetry CDW phase. Our results reveal a competition between the $2\times2\times2$ and $2\times2\times4$ stacking phases, explaining the ongoing debate in the community, as both structures are dynamically stable and almost degenerate in free energy. 
The atomistic characterization of the CDW phase allows us to study its superconductivity and determine that it is phonon-mediated.

\begin{figure*}
\includegraphics[width=\textwidth]{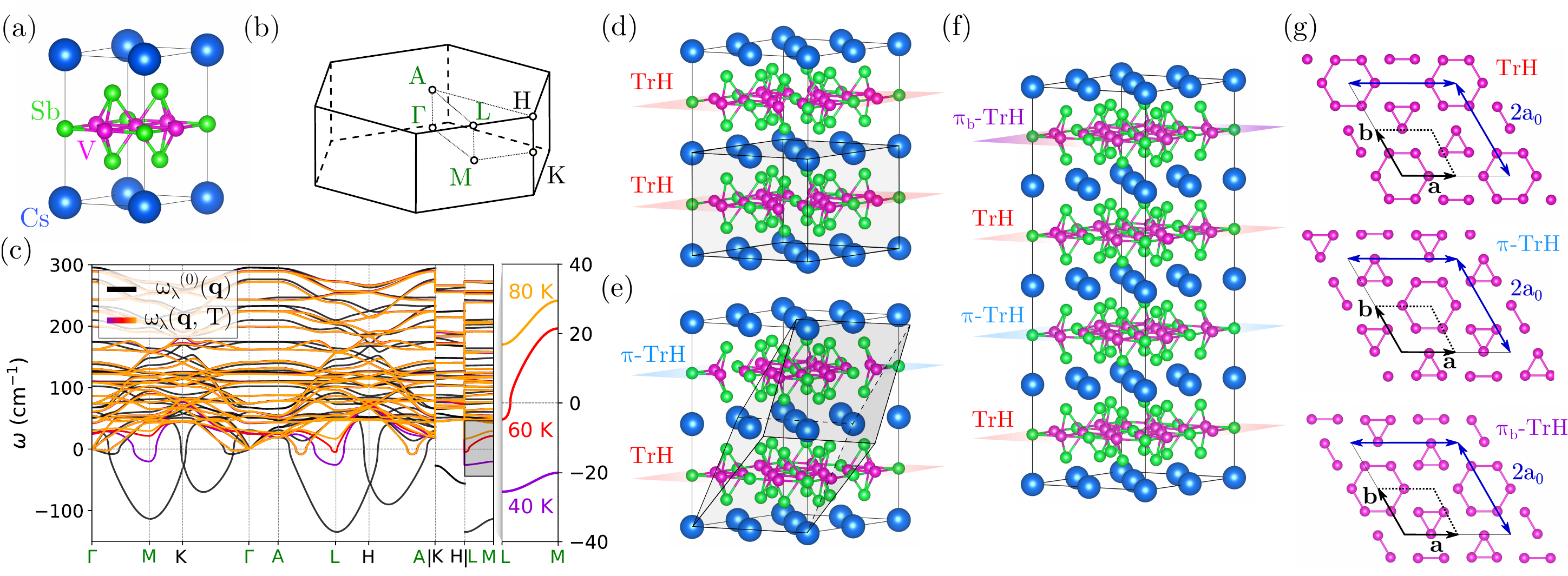}% Here is how to import EPS art
\caption{\label{fig:Fig1} Structural and dynamical properties of CsV$_3$Sb$_5$. (a) Atomic structure of the pristine high-symmetry phase. (b) Visualization of the Brillouin zone of the high-symmetry phase and its high-symmetry points. (c) Harmonic and SSCHA Hessian phonon spectra of the high-symmetry phase calculated
at 40, 60 and 80 K. Imaginary modes are depicted with negative values. The high symmetry wave vectors calculated explicitly and  later used for the Fourier interpolation are highlighted in green. (d),(e),(f) Atomic structure of the TrH, $\pi$-TrH and 4TrH phases. In the case of TrH and $\pi$-TrH, their respective primitive cells are highlighted. (g) Visualization of the out-of-plane shift of the vanadium TrH layers that distinguish the emerging low-symmetry metastable phases.}
\end{figure*}

Above $T_{CDW}$ the primitive structure of CsV$_3$Sb$_5$ belongs to the P6/mmm (No. 191) space-group. It consists of a Cs hexagonal unit-cell, inside which a Sb hexagonal- and a V Kagome-layer are sandwiched between Sb honeycomb-layers (see Fig. \ref{fig:Fig1}(a)). As shown in Fig. \ref{fig:Fig1}(c), at the harmonic level, this phase exhibits lattice instabilities at the zone-border wave vectors $\mathbf{q}_M$=(1/2,0,0) and $\mathbf{q}_L$=(1/2,0,1/2)~\cite{subediHexagonaltobasecenteredorthorhombic4QCharge2022, siChargeDensityWave2022} (we express $\mathbf{q}$ vectors in units of the reciprocal space lattice vectors). This means that
\begin{equation}[\omega^{(0)}_\lambda(\mathbf{q})]^2=\frac{\partial^2V(\mathbf{R})}{\partial Q^2_\lambda(\mathbf{q})}<0
%\,\mathrm{\,for\,}\vec{q}=\vec{q}_L\,\mathrm{and}\,\vec{q}_M,
\label{eq:harmonic_w}
\end{equation}
both for the lowest energy modes at $\mathbf{q}_L$ and $\mathbf{q}_M$, where $\omega^{(0)}_\lambda(\mathbf{q})$ is the harmonic frequency of mode $\lambda$, $V(\mathbf{R})$ is the BO energy, and $Q_\lambda(\mathbf{q})$ the order-parameter related to the ionic displacement described by the mode $\lambda$ with wave vector $\mathbf{q}$ (see Supplementary Material). 
Harmonic imaginary phonon frequencies thus signal that a structure is not a minimum of the BO energy surface, but that it lowers $V(\mathbf{R})$ along the lattice distortion described by the order parameter.

The P6/mmm  high-symmetry phase must be, however, the free energy minimum above $T_{CDW}$. This clearly illustrates the breakdown of any approach based on the BO energy to assign the thermodynamic stability of any CDW prototype, which has been the case so far~\cite{TanChargeDensityWaves2021,ratcliffCoherentPhononSpectroscopy2021,subediHexagonaltobasecenteredorthorhombic4QCharge2022a}, and the need to consider ionic kinetic effects as well as anharmonicity at a nonperturbative level to estimate the phase diagram of CsV$_3$Sb$_5$.
The SSCHA is a perfectly suited method for that, as it variationally minimizes the total free energy of the system calculated with a trial density matrix $\rhos$,
\begin{equation}
    F[\rhos] = \langle K+V(\mathbf{R})\rangle_{\rhos}-TS[\rhos]\,,
\label{eq:f}
\end{equation}
fully considering the BO potential as well as the ionic kinetic energy $K$ and entropy $S$. The variational parameters are the centroid positions $\Rc$, which determine at the end of the minimization the most probable ionic positions, and the auxiliary force constants $\Phit$, related to the broadening of the ionic probability distribution function around the centroid positions.
In analogy to the harmonic case (see Eq. \eqref{eq:harmonic_w}), an imaginary phonon frequency obtained from the Hessian of the SSCHA free energy,
\begin{equation}
[\omega_\lambda(\mathbf{q},T)]^2=\frac{\partial^2F[\rhos]}{\partial Q^2_\lambda(\mathbf{q})}\,,
\label{eq:f_w}
\end{equation}
signals that a structure is unstable towards the distortion described by a given normal mode. The $\omega_\lambda(\mathbf{q},T)$ phonon frequencies, usually called SSCHA Hessian phonons, are temperature-dependent, include nonperturbative anharmonicity, and can be understood as the position of the peaks of the phonon spectral function in the static limit~\cite{BiancoSecondOrderStructural2017,MonacelliTime-DependentSelf2021,LihmGaussianTime-Dependent2021}.

The temperature-dependent anharmonic spectra obtained from the Hessian phonons is shown in Fig. \ref{fig:Fig1}(c) for the high-symmetry phase. 
In agreement with previous findings~\cite{gutierrez-amigoPhononCollapseAnharmonic2024}, at temperatures above 65 K this phase becomes dynamically stable thanks to ionic entropy, i.e. it becomes a local minimum of the free energy.
The phonon instability at L is the last to disappear. In a second-order phase transition scenario, one could argue that the CDW phase must adopt a $2\times2\times2$ modulation below 65 K as a direct consequence of the phonon instability at L. 
However, given the contradictory findings regarding the out-of-plane modulation and the most probable first-order character of the transition, such an argument appears overly simplistic. Notably, the whole LM phonon branch becomes unstable at temperatures close to the transition. This is an indication that $2\times2\times X$ modulated phases with lower free-energy than the high-symmetry phase may be present, which can not be ignored as possible charge-reconstructed phases. The possibility that any ordered phase has a lower free energy than the P6/mmm one above 65 K cannot be discarded, which would be the case in a first-order phase transition.

Solving the CDW phase diagram at low temperatures, therefore, requires to calculate the free energy of model CDW distorted phases as a function of temperature.
The lowest-energy phonon branch in the LM high-symmetry line predominantly corresponds to an in-plane distortion of the vanadium Kagome layer. At the L and M high-symmetry points, the distortion is commensurate with a $2\times2\times2$ supercell, and the lowest energy modes transform under the $L_2^-$ and $M_1^+$ irreducible representations, respectively. To explore all possible emergent phases, we investigate the local-minima of the BO energy surface by relaxing independently distorted lattice configurations along the $L_2^-$ and $M_1^+$ directions. This process reveals six local-minima of $V(\mathbf{R})$ (see Supplementary Material) with a lower BO energy than the high-symmetry phase, offering plausible models of the CDW reconstruction. Among them, both SoD and TrH arrangements of the vanadium layer are present.
Remarkably, none of these models is dynamically stable at the harmonic level, further corroborating the necessity of including anharmonicity at a nonperturbative level to describe not only the high-symmetry phase of CsV$_3$Sb$_5$, also its CDW phase.

\begin{figure}
\includegraphics[width=0.5\textwidth]{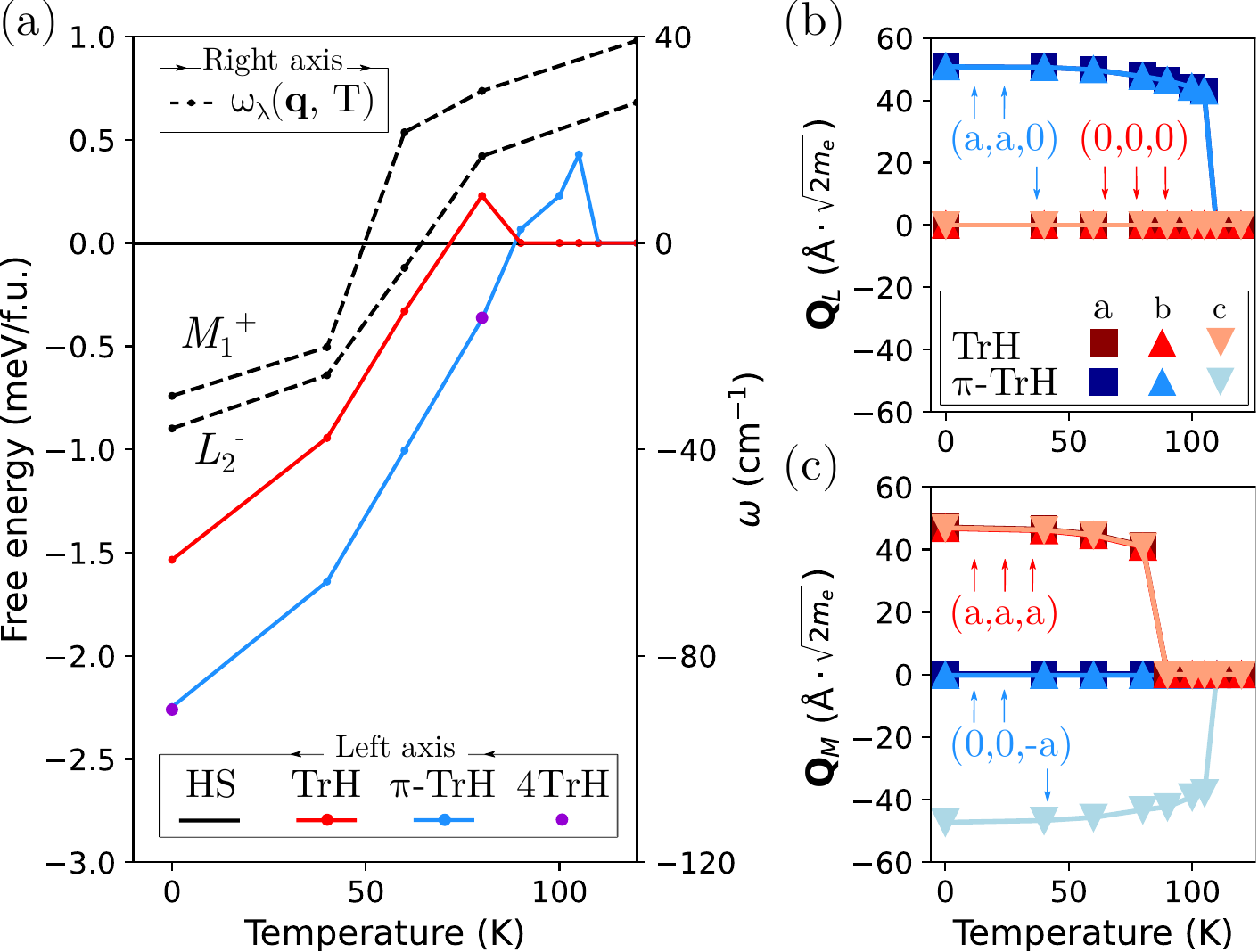}% Here is how to import EPS art
\caption{\label{fig:Fig2}Energetics and structural evolution as a function of temperature of the TrH stacking orders. (a) Free energy as a function of temperature of the possible different CDW phases (continuous lines), compared to the phonon collapse of the high symmetry phase (dashed lines). The free energy is computed in the warming up process, while frequencies in the cooling. Projection of TrH and $\pi$-TrH CDW crystal structures as a function of temperature onto the order parameters of $L_2^-$ (b) and $M_1^+$ (c) modes. Considering that these modes are non-degenerate at a given wave vector, but the star is composed of three wave vectors, the order parameter is represented as $\mathbf{Q}_{L,M}=(a,b,c)_{L,M}$, where $a,b,c$ represent the projection onto each of the vectors in the star. 
}
\end{figure}

In a second step, we relax these model CDW phases considering the ionic kinetic energy and anharmonicity within the SSCHA at 0 K, without imposing any symmetry constraints. In order to broaden the analysis, we also consider a new configuration obtained by an arbitrary distortion at the mid-point of the LM branch. This structure, with a $2\times2\times4$ modulation, lacks any symmetry (P1 (No. 1) space-group). The energy landscape obtained with Eq. \eqref{eq:f} simplifies to just three local minima. 
In fact, all charge ordering patterns apart from the TrH, including the SoD, are thermodynamically unstable and transition into a TrH ordering. Indeed, the TrH reconstruction is found in three different phases distinguished by the stacking-order (see Fig. \ref{fig:Fig1}(d-f)): the TrH, $\pi$-TrH, and 4TrH phases. 
The TrH phase belongs to the P6/mmm (No. 191) space-group given that the charge is reconstructed without any out of plane modulation, preserving the hexagonal symmetry, and it is energetically favored by 1.53 meV/f.u with respect to the high-symmetry phase (see Fig. \ref{fig:Fig2}(a)). 
The $\pi$-TrH phase belongs to the Fmmm (No. 69) space-group, and shows an ABABA stacking order commensurate in a $2\times2\times2$ supercell that breaks hexagonal symmetry. It is favored with respect to the high-symmetry phase by 2.25 meV/f.u. The 4TrH phase is practically degenerate in energy with the $\pi$-TrH, as it is reconstructed into a $2\times2\times4$ modulated ABACA stacking of $\pi$ shifted TrH layers. 
However, this phase belongs to the P1 (No. 1) space-group, as  no symmetry is gained in the relaxation, although it is not far from the Cccm (No. 66) space group.
While the TrH phase is metastable, we can conclude that the $\pi$-TrH and the 4TrH phases are the ground state CDW phases of the system. External perturbations such as strain or impurities may influence the presence of one or the other, naturally explaining the controversial presence of $2\times2\times2$ and $2\times2\times4$ orders in experiments~\cite{zhaoCascadeCorrelatedElectron2021, kangChargeOrderLandscape2023, ratcliffCoherentPhononSpectroscopy2021, ningDynamicalDecodingCompetition2024, liObservationUnconventionalCharge2021, subiresOrderdisorderChargeDensity2023,broylesEffectInterlayerOrdering2022, ortizFermiSurfaceMapping2021,jinPhaseInterlayerShift2024, xiaoCoexistenceMultipleStacking2023, kautzschStructuralEvolutionKagome2023, stahlTemperaturedrivenReorganizationElectronic2022,jinPhaseInterlayerShift2024}, and supporting the possibility of having disordered domains of different stacking. 

As temperature increases, the free energy difference of the CDW candidates with respect to the pristine phase is reduced. At 89 K the latter phase becomes the ground state, in perfect agreement with the experimentally observed $T_{CDW}=94$ K value~\cite{frachetColossalAxisResponse2024}. Remarkably, the $\pi$-TrH ordering persists up to 105 K, where it transitions abruptly into the high-symmetry phase (Fig. \ref{fig:Fig2} (b-c)). Within this temperature range, while the high-symmetry phase is thermodynamically favored, the $\pi$-TrH and the 4TrH phases remain metastable. Upon cooling from ambient conditions, a similar phenomena occurs. The high-symmetry phase is energetically favored until 89 K, after which the $\pi$-TrH and 4TrH phases become the ground state. However, down to temperatures around 65 K, the high-symmetry phase is dynamically stable given that the frequency of the $L_2^-$ mode is still positive. Thus, our calculations reproduce the first-order character of the CDW phase transition, in agreement with experiments~\cite{frachetColossalAxisResponse2024}. Even if it will be difficult to realize it experimentally given the small energy differences, our results allow a hysteresis region of almost 40 K.

\begin{figure*}
\includegraphics[width=\textwidth]{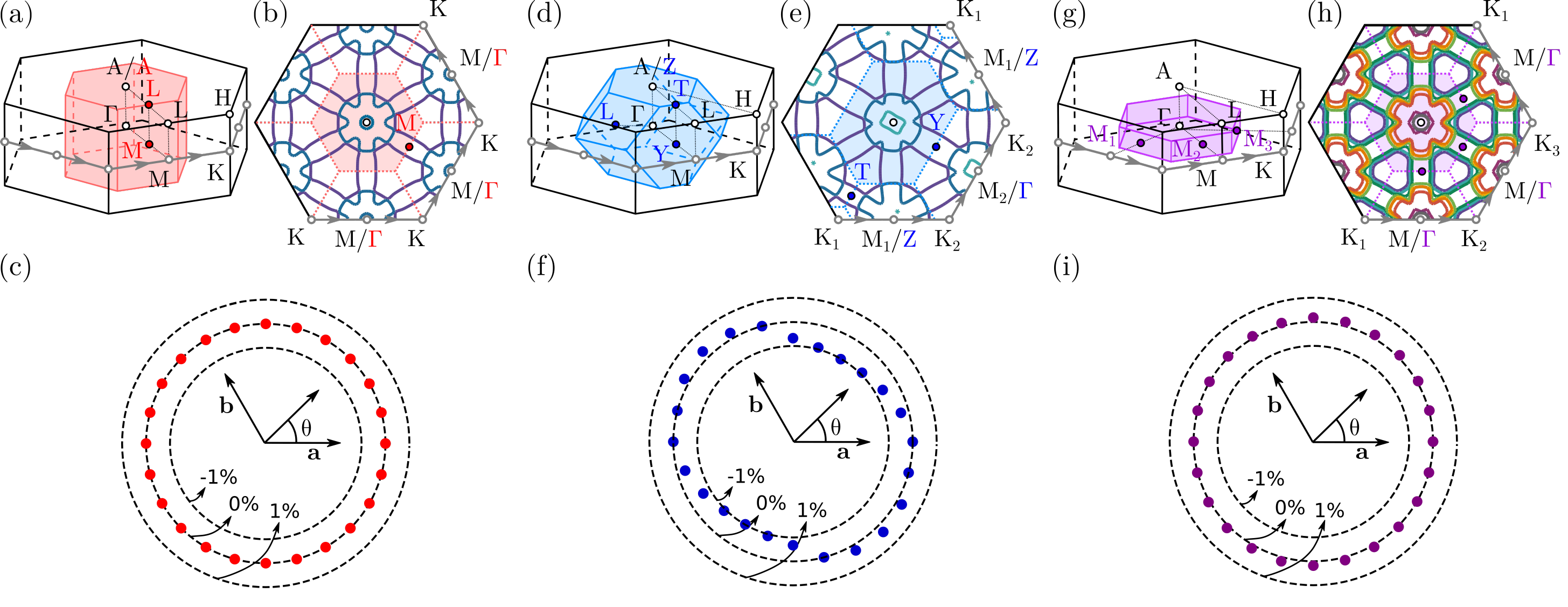}% Here is how to import EPS art
\caption{\label{fig:Fig3} Fermiology and direction dependent conductivity of the low temperature CDW stable and metastable phases of CsV$_3$Sb$_5$. (a) Illustration of the Brillouin zone of the TrH phase. High-symmetry points of the primitive Brillouin zone are highlighted in color, while those of the pristine high-symmetry phase in black and gray. (b) Cut of the Fermi surface at $k_z=0$ plane of the TrH phase visualized in the Brillouin zone of the high-symmetry phase. The smaller Brillouin zone of the TrH phase is shadowed in color. (c) Direction dependent in-plane anisotropy of the electrical conductivity $\sigma$ for the TrH phase. The anisotropy is defined as $1-\sigma(\theta)/\sigma(\theta=0)$, and is represented as a function of an angle $\theta$ defined by the transport direction and the lattice-vector $\mathbf{a}$. (d-f) and (g-i) same as (a-c) for the $\pi$-TrH and 4TrH phases.}
\end{figure*}

Despite our calculations show that the CDW ground state structures, the $\pi$-TrH and 4TrH phases, have a clear broken hexagonal symmetry due to the out-of-plane stacking, transport experiments show no intrinsic anisotropy in the electrical conductivity~\cite{guoCorrelatedOrderTipping2024}. In order to solve this apparent contradiction, we study the fermiology of these phases and compare it with the symmetric metastable TrH, which consequently keeps a Fermi surface with hexagonal symmetry (see Fig. \ref{fig:Fig3}(b)).

The $\pi$-TrH corresponds to a face centered orthorhombic structure, and the now distinct M points (M$_1$ and M$_2$) have  a clearly different fermiology, breaking the hexagonal symmetry (see Fig. \ref{fig:Fig3}(e)): M$_2$ shows a Fermi pocket absent at M$_1$.
In the 4TrH phase, despite the absence of six-fold symmetry, the Brillouin zone is quasi-hexagonal and the Fermi surface at a first glance appears to respect six-fold symmetry, even if three different M points exist 
%in the zone borders of the low-symmetry Brillouin zone 
(see Fig. \ref{fig:Fig3}(g)).
By making use of semi-classical Boltzmann theory, we analyze the impact of the different fermiology on the anisotropy of the electrical conductivity. 
%We employed these calculations on the in-plane electronic bands to estimate anisotropy in conductivity based on semi-classical Boltzmann theory. 
Figs. \ref{fig:Fig3}(c,f,i) show the angular dependence of the anisotropy. While the TrH phase shows symmetry-imposed isotropic conductivity, the $\pi$-TrH phase and the 4TrH phases do not. The $\pi$-TrH phase, shows a -1\% of anisotropy that peaks along the $\mathbf{a}+\mathbf{b}$ direction. This small effect, is further reduced by the out-of-plane disorder, as the 4TrH phase shows anisotropy below -0.2\%. 
These results demonstrate that, thanks to the different stacking of the TrH layers, the breaking of the six-fold symmetry is perfectly compatible with an isotropic conductivity, consistent with experiments~\cite{guoCorrelatedOrderTipping2024}. 
The out-of-plane disorder is also compatible with the electronic band splitting observed in ARPES~\cite{kangChargeOrderLandscape2023} and the phonon band splitting observed with ultrafast time-resolved reflectivity measurements around 1.3 THz ~\cite{liangCoherentPhonon2025} (see Supplementary Material), which were before attributed to a possible TrH-SoD stacking that turns out clearly thermodynamically unstable in our calculations. 

\begin{figure*}
\includegraphics[width=\textwidth]{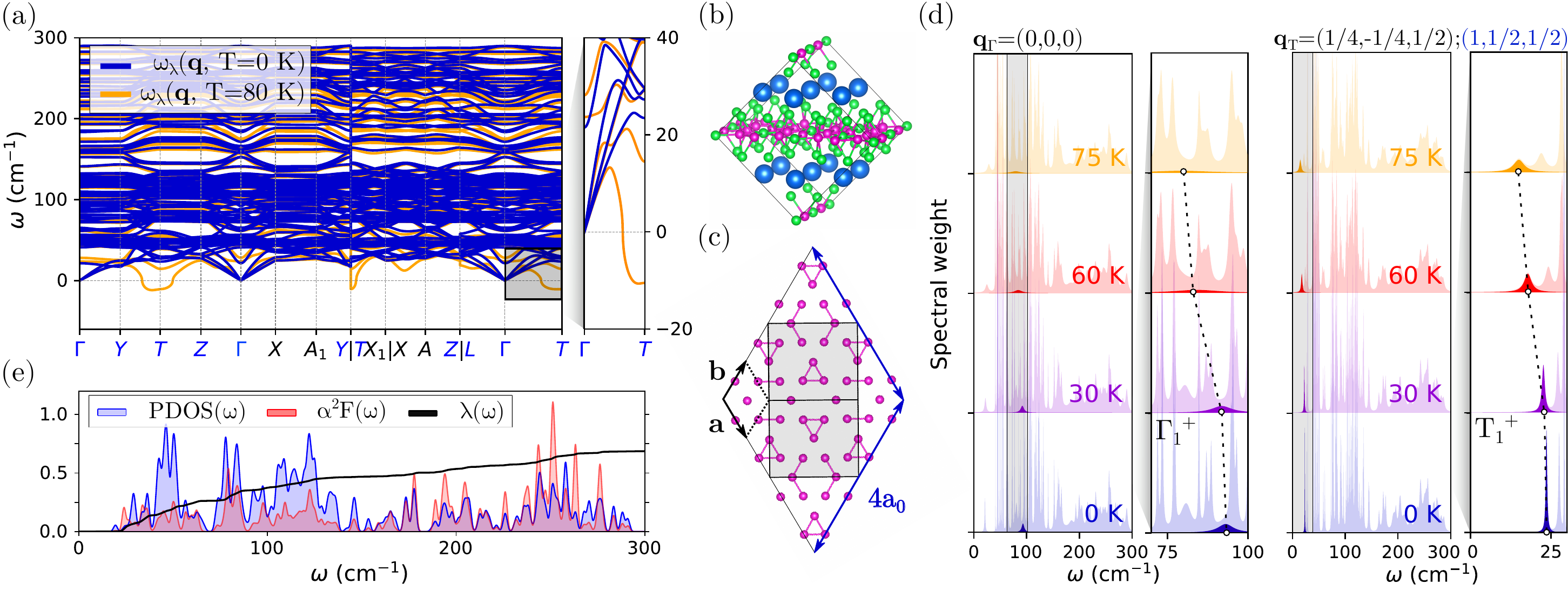}% Here is how to import EPS art
\caption{\label{fig:Fig4} Renormalization of the phonon-spectra of the $\pi$-TrH phase with temperature. (a) Phonon spectra computed by a Fourier interpolation from a $2\times2\times2$ $\mathbf{q}$-grid of the primitive lattice of the $\pi$-TrH phase. The high symmetry wave vectors calculated explicitly and  later used for the Fourier interpolation are highlighted in blue. (b) Atomic structure obtained from the distortion of the $T_1^+$ lattice-instability. (c) Top-view of the reconstruction of the TrH vanadium layer obtained from the distortion of the $T_1^+$ lattice-instability. The primitive unit-cell is highlighted. (d) Anharmonic spectral function of the $\pi$-TrH phase including phonon-phonon and electron-phonon interactions. In the left panel the spectral function at $\Gamma$ is represented, where the contribution of the $\Gamma_1^+$ mode is highlighted. In the right panel the spectral function is computed at $T$, highlighting the $T_1^+$ mode. (e) Phonon density of states PDOS($\omega$), along with the Eliashberg function $\alpha^2F(\omega$), and the integrated electron-phonon parameter $\lambda(\omega$) of the $\pi$-TrH phase.}
\end{figure*}

The fact that the CDW ground state structures are consistent with transport and ARPES data support their validity. Nevertheless, their dynamical stability needs to be confirmed by calculating its SSCHA Hessian phonons in a larger grid of their Brillouin zone. 
Such anharmonic temperature-dependent phonon spectra is plotted in Fig. \ref{fig:Fig4}(a) for the $\pi$-TrH phase. 
%Due to computational limitations, the 4TrH phase could not be considered.
Starting from 0 K, the structure remains dynamically stable up to temperatures close to the CDW transition. The absence of lattice instabilities confirms that the CDW persists up to $T_{CDW}$, and, hence, if an order-disorder transition occurs from the $\pi$-TrH to the 4TrH phase, it must be of first-order character, consistent with experimental reports~\cite{jinPhaseInterlayerShift2024}.
Interestingly, a lattice-instability appears at approximately 80 K at the T point ($\mathbf{q}=(1/4,-1/4,1/2)$ in units of the reciprocal lattice vectors of the high-symmetry phase or (1,1/2,1/2) in units of the reciprocal lattice vectors of the $\pi$-TrH phase), rather than at $\Gamma$. This phonon-mode, which transforms under the irreducible representation $T_1^+$, is non-degenerate. The resulting distorted structure is commensurate in a $4\times4\times2$ supercell of the high-symmetry phase, and belongs to the Cmmm (No. 65) space-group (see Fig. \ref{fig:Fig4}(b,c)). 
Even if this instability could signal the presence of another phase between 80 and 89 K, which does not seem to be observed experimentally~\cite{frachetColossalAxisResponse2024}, it could be related to a second CDW phase with a larger in-plane order that has been recently observed under compression~\cite{zhengEmergentChargeOrder2022, stierPressureDependentElectronicSuperlattice2024}. In fact, the 
$T_1^+$ mode shows a strong renormalization with isotropic pressure in our calculations, with a tendency to become more unstable (see Supplementary Material), consistent with experiments.

By calculating the phonon spectral function with the dynamical extension of the SSCHA theory~\cite{BiancoSecondOrderStructural2017,MonacelliTime-DependentSelf2021,LihmGaussianTime-Dependent2021} along with first-principle electron-phonon calculations, we predict that the softening of the $T_1^+$ mode should be observable by, for instance, inelastic X-ray experiments. As shown in Fig. \ref{fig:Fig4}(d), this mode undergoes a significant renormalization before collapsing at 80 K, but it remains separated  from the rest of the spectra, allowing its spectral peak to remain well-defined still at 75 K. 
We perform a similar analysis for the $\Gamma_1^+$ phonon mode, the one that emerges as the condensation of the $L_2^-$ mode of the high-symmetry phase. This mode exhibits a marked red-shift from 93 to 78 cm$^{-1}$ between 0 and 75 K, accompanied by an increase in broadening, which makes it practically vanish from the spectrum at 75 K. This behavior aligns well with previous Raman measurements, which report a strong renormalization of a $\Gamma_1^+$ mode at similar energies~\cite{heAnharmonicStrongcouplingEffects2024}.

Having determined the charge-ordered structure allows us to study the origin of the intriguing superconductivity of CsV$_3$Sb$_5$ from first principles. Experimentally, the superconducting critical temperature ($T_c$) of approximately 2.5 K~\cite{OrtizCsV3Sb5Superconductivity} increases under pressure displaying a double-dome structure before superconductivity is ultimately suppressed~\cite{chenDoubleSuperconductingDome2021, yu_unusual_2021}.
The first of these two $T_c$ maxima coincides with the previously discussed emergence of the in-plane superlattice phase, while the second corresponds to the melting of the CDW, where $T_c$ increases approximately up to 8 K~\cite{zhengEmergentChargeOrder2022}. 
Despite the electron-phonon interaction was estimated to be too weak to explain the observed $T_c$~\cite{TanChargeDensityWaves2021}, pointing to an unconventional mechanism~\cite{wuNatureUnconventionalPairing2021,TanChargeDensityWaves2021,PhysRevB.111.134507}, the phonon mediated scenario cannot be ruled out since no calculation has been performed thus far in the CDW phase considering the crucial role of anharmonicity. 
By combining the SSCHA anharmonic phonon spectrum of the $\pi$-TrH phase with \emph{ab initio} electron-phonon matrix elements (see Supplementary Material), we calculate a $T_c$ of 2.7 K for the CDW phase, in good agreement with experiments. 
As shown in Fig. \ref{fig:Fig4}(e), the contribution to the electron-phonon coupling constant $\lambda$ is rather evenly distributed among all phonon modes.
In order to test the impact of the CDW on the superconducting critical temperature, we repeat the calculation for the high-symmetry phase assuming that its phonon spectra is the anharmonic one at 100 K. Interestingly, $T_c$ is enhanced to 12.1 K, which is consistent with the large  increase of the critical temperature observed experimentally when the CDW is melted under pressure. In fact, the CDW strongly suppresses the electron-phonon coupling constant (see Suppementary Material).
Our results demonstrate that superconductivity in CsV$_3$Sb$_5$ is conventional, in the sense that it is mediated by the electron-phonon coupling, and that there is a large interaction between superconductivity and the CDW reconstruction.

In summary, by performing first-principles calculations including the ionic kinetic energy, entropy, and anharmonicity, we resolve the phase diagram of CsV$_3$Sb$_5$ and its CDW phase. Our results yield a first-order CDW phase transition at 89 K, in good agreement with experimental results. The obtained CDW ground-state ordering is inherently three dimensional, with the nature of the stacking being hard to resolve. 
However, being consistent with transport and Raman data, the obtained $\pi$-TrH and 4TrH phases are realistic models of the CDW. 
With the CDW solved and having a realistic description of its phonon spectrum, we determine that superconductivity is mediated by phonons and that the CDW strongly suppresses its critical temperature.

\begin{acknowledgments}
We thank Fernando de Juan for insightful discussions. This project was supported by the
PID2022-142861NA-I00 and PID2022-142008NB-I00 projects funded by MICIU/AEI/10.13039/501100011033 and FEDER, UE;
the Simons Foundation through the Collaboration on New Frontiers in Superconductivity (Grant No. SFI-MPS-NFS-00006741-10); the Department of Education, Universities and Research of the Eusko Jaurlaritza and
the University of the Basque Country UPV/EHU (Grant
No. IT1527-22);
Canada Excellence Research Chairs
Program for Topological Quantum Matter;
NSERC Quantum Alliance France-Canada; and Diputación Foral de Gipuzkoa Programa Mujeres y Ciencia
M.A. acknowledges a PhD scholarship from the Materials Physics Center.
M.G.-A. was supported by Jane and Aatos Erkko
Foundation, Keele Foundation, and Magnus Ehrnrooth
Foundation as part of the SuperC collaboration. 
The authors acknowledge the technical and human support provided by the DIPC Supercomputing Center.
\end{acknowledgments}

\bibliography{CsV3Sb5_paper}% Produces the bibliography via BibTeX.

\end{document}

\renewcommand{\thefigure}{S\arabic{figure}}
\renewcommand{\thetable}{S\arabic{table}}

\newcommand{\ion}[1]{\textcolor{teal}{(\textbf{Ion:} #1)}}
\newcommand{\Rc}{\boldsymbol{\mathcal{R}}}
\newcommand{\Phit}{\boldsymbol{\Phi}}
\newcommand{\rhos}{\rho_{\Rc,\Phit}}

\preprint{APS/123-QED}

%\title{Supplementary Material: \\ Symmetry-broken charge-ordered ground state in CsV$_3$Sb$_5$ Kagome metal}% Force line breaks with \\
\title{Supplementary Material: \\ Symmetry-broken ground state and phonon mediated superconductivity in Kagome CsV$_3$Sb$_5$}

\author{Manex Alkorta$^\mathrm{1,2}$}
\author{Martin Gutierrez-Amigo$^\mathrm{3}$}
\author{\DJ or\dj e Dangi\'c$^\mathrm{1,2}$}
\author{Chunyu Guo$^\mathrm{4}$}
\author{Philip J. W. Moll$^\mathrm{4}$}
\author{Maia G. Vergniory$^\mathrm{5,6,7}$}
\author{Ion Errea$^\mathrm{1,2,5}$}
 \affiliation{$^1$Centro de Física de Materiales (CFM-MPC), CSIC-UPV/EHU, Donostia, 20018, Spain}
 \affiliation{$^2$Department of Applied Physics, University of the Basque Country (UPV/EHU), Donostia, 20018, Spain}
 %\altaffiliation[Also at ]{Physics Department, XYZ University.}%Lines break automatically or can be forced with \\}%
\affiliation{$^3$
Department of Applied Physics, Aalto University School of Science, FI-00076 Aalto, Finland}%
\affiliation{$^4$Max Planck Institute for the Structure and Dynamics of Matter, Hamburg, 22761, Germany}
 \affiliation{$^5$Donostia International Physics Center (DIPC), Donostia, 20018, Spain}
\affiliation{$^6$Département de Physique et Institut Quantique, Université de Sherbrooke, Sherbrooke, J1K 2R1 Québec, Canada.}
\affiliation{$^7$Regroupement Qu\'eb\'ecois sur les Mat\'eriaux de Pointe (RQMP), Quebec H3T 3J7, Canada}

\date{\today}% It is always \today, today,

\maketitle

\section{Computational approach}

For the analysis of the Born Oppenheimer (BO) energy landscape and its local minima, structural relaxations were performed within the density functional theory (DFT) framework using the Vienna Ab Initio Simulation Package (VASP) \cite{kresseEfficientIterativeSchemes1996}. A plane-wave energy cutoff of 400 eV was used, along with a Methfessel-Paxton smearing of 0.1 eV. The Brillouin zone was sampled using a $\Gamma$-centered $16\times16\times8$ $\mathbf{k}$-point mesh for the high-symmetry phase, proportionally reduced for the low-symmetry supercells. Projector augmented-wave (PAW) pseudopotentials were employed with valence configurations of 5s$^2$5p$^6$6s$^1$ for Cs, 3p$^6$3d$^4$4s$^1$ for V, and 5s$^2$5p$^3$ for Sb. To capture the Van der Waals interaction along the out-of-plane direction, the optB88-vdW exchange-correlation functional was used \cite{klimesChemicalAccuracyVan2009}, which describes accurately the lattice-parameters of CsV$_3$Sb$_5$ (see Table \ref{tab:Table1}).
\begin{table}
\caption{\label{tab:Table1} Lattice vectors of the high-symmetry phase compared to experimental observations.}
\begin{ruledtabular}
\begin{tabular}{cccc}
 {}&{$a$} (\AA)&{$b$} (\AA)&{$c$} (\AA)\\\hline
 Experimental value from \cite{ortizNewKagomePrototype2019} & 5.4949 & 5.4949 & 9.3085 \\
 DFT with optB88-vdW & 5.5058 & 5.5058 & 9.3308 \\
\end{tabular}
\end{ruledtabular}
\end{table}

\begin{figure}
\includegraphics[width=0.5\textwidth]{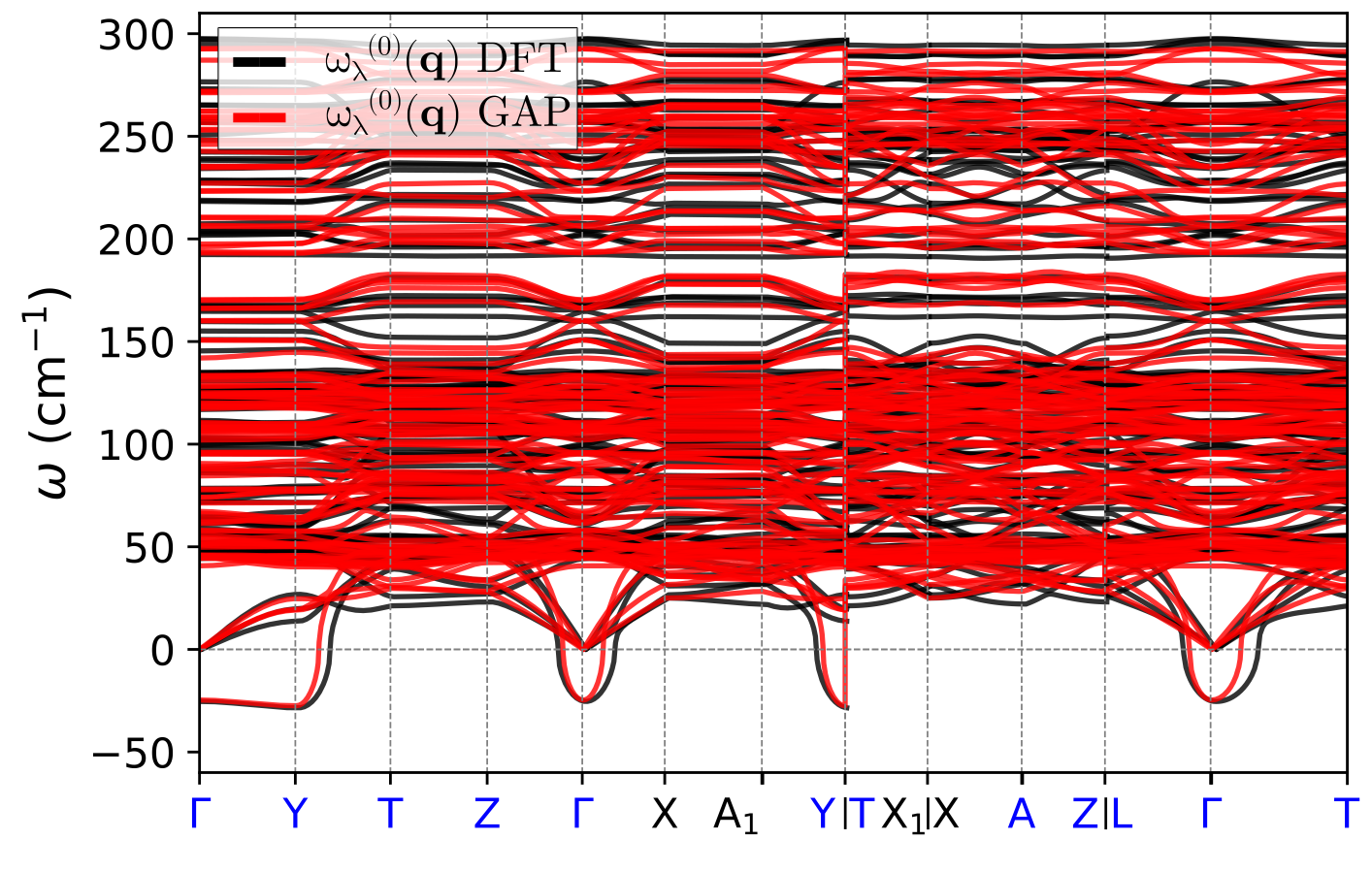}% Here is how to import EPS art
\caption{\label{fig:S1} Benchmark on harmonic phonons of the $\pi$-TrH phase calculated in a $2\times2\times2$ supercell. The high-symmetry wave vectors calculated explicitly, and  later used for the Fourier interpolation, are highlighted in blue.}
\end{figure}

Ionic quantum and anharmonic effects, as well as the phonon-phonon interaction, were included via the stochastic self-consistent harmonic approximation (SSCHA) \cite{monacelliStochasticSelfconsistentHarmonic2021}. Given the size of the supercells considered, energies, forces and stresses needed for the SSCHA minimizations were computed using an iteratively trained gaussian approximation potential (GAP) \cite{bartokGaussianApproximationPotentials2010}. The dataset is composed of over 5000 configurations, which were generated with the SSCHA probability distribution function and calculated with DFT, using the setup described in the previous paragraph. This offers physical ionic distribution functions and, consequently, meaningful configurations. From the whole dataset, 1100 configurations were iteratively selected to train the final potential. The resulting GAP model achieves root-mean-square errors (RMSE) of 0.13 meV/atom for energies, 21.27 meV/$\mathrm{\AA}$ for forces and 0.13 meV/\AA$^3$ for stresses, evaluated in the full dataset. To validate the accuracy of the interatomic potential, a benchmark on harmonic phonons of the $\pi$-TrH phase, $\omega^{(0)}_\lambda(\mathbf{q})$, was performed. The result confirms that the GAP potential accurately reproduces the Born-Oppenheimer energy surface, $V(\mathbf{R})$, and its derivatives, as shown in Fig. \ref{fig:S1}.
The harmonic phonon-spectra, $\omega^{(0)}_\lambda(\mathbf{q})$, were computed using the finite displacement method. For DFT based phonons, we employed the PHONOPY code \cite{phonopy-phono3py-JPCM, phonopy-phono3py-JPSJ} interfaced with VASP, whereas ASE \cite{HjorthLarsen_2017} was used together with the GAP model. The temperature-dependent Hessian phonon spectra, $\omega_\lambda(\mathbf{q}, T)$, were calculated using the SSCHA method, employing the GAP to evaluate forces. For the high-symmetry phase, Hessian phonon spectra were computed without any approximation. In the case of the $\pi$-TrH phase, the bubble approximation was employed \cite{BiancoSecondOrderStructural2017}.

The in-plane electrical conductivity was calculated using the semi-classical Boltzmann transport theory \cite{scheidemantelTransportCoefficientsFirstprinciples2003}, implemented in the BoltzTrap2 code \cite{madsenBoltzTraP2ProgramInterpolating2018}. Within this framework, the electrical conductivity tensor $\boldsymbol{\sigma}$ is given by
\begin{equation}
    \sigma^{\alpha,\beta}(T,\mu) = \sum_{n,\mathbf{k}}\frac{e^2}{\Omega N_\mathbf{k}}\left[-\frac{\partial f(T, \mu, \varepsilon_{n,\mathbf{k}})}{\partial{\varepsilon}}\right]\tau_{n,\mathbf{k}}v^\alpha_{n,\mathbf{k}}v^{\beta}_{n,\mathbf{k}}\,,
\end{equation}
where $\alpha,\beta$ are Cartesian-coordinates, $e$ is the electron charge, $\Omega$ the unit-cell volume, and $N_\mathbf{k}$ is the number of $\mathbf{k}$-points in the sum. $f(T,\mu,\varepsilon_{n,\mathbf{k}})$ is the Fermi-Dirac distribution function at temperature $T$ and chemical-potential $\mu$. $\varepsilon_{n,\mathbf{k}}$, $\mathbf{v}_{n,\mathbf{k}}$ and $\tau_{n,\mathbf{k}}$ are the energy, group velocity and electronic relaxation time of the electron-band with index $n$ at the wave vector $\mathbf{k}$. The latter is in practice assumed constant $\tau_{n,\mathbf{k}}\approx\tau$ for all bands and $\mathbf{k}$-points. We evaluated the conductivity at $T=10$ K and $\mu=\varepsilon_F$. To precisely include features along the high-symmetry lines of the hexagonal Brillouin zone at $k_z=0$, the electronic bands were calculated in their respective commensurate $2\times2\times X$ supercells, using a $16\times16\times1$ $\mathbf{k}$-mesh. 

\section{Relaxation of distorted geometries}

To account for all possible charge reconstructions, the pristine phase of CsV$_3$Sb$_5$ was distorted along the directions corresponding to the harmonic dynamical instabilities identified at a DFT level: the $L_2^-$ and the $M_1^+$ modes. While these modes are non-degenerate, hexagonal symmetry of the lattice leads to threefold degeneracy in wave vector space, with each instability characterized by a star of three symmetry equivalent wave vectors $\mathbf{q}^{(s)}$ (with $s=1,2,3$). Therefore, arbitrary distortions along these modes, must be constructed as linear combinations of the components of the star. For instance, considering the $\lambda\equiv L_2^-$ mode, and denoting the wave vectors of the star as $\mathbf{q}_L^{(s)}$, an arbitrary distortion relative to the high-symmetry structure is given by
\begin{equation}
\Delta \mathbf{R}_{i,a} = \frac{1}{\sqrt{M_i}}\sum_{\mathrm{s=1}}^3Q_{\lambda}(\mathbf{q}_L^{(s)})\boldsymbol{\epsilon}_{\lambda,i}(\mathbf{q}_L^{(s)})\cdot e^{i\mathbf{q}^{(s)}\cdot \mathbf{T}_a}\,,
\label{eq:dR}
\end{equation}
where $i$ is an atomic index at the reference (primitive) unit cell, and $\mathbf{T}_a$ denotes a lattice translation into the unit cell labeled as $a$. $M_i$ is the mass of the ion $i$, and $Q_{\lambda}$ is the order parameter (amplitude) associated to the distortion along the $\boldsymbol{\epsilon}_{\lambda}$ polarization vector. In fact, the order parameter can be compactly written as a three component vector, $\mathbf{Q}_L=(Q_\lambda(\mathbf{q}_L^{(1)}),Q_\lambda(\mathbf{q}_L^{(2)}),Q_\lambda(\mathbf{q}_L^{(3)}))\equiv(a,b,c)_L$, which describes a distortion in the subspace defined by the polarization vector basis $\{\boldsymbol{\epsilon}_\lambda(\mathbf{q}_L^{(1)}), \boldsymbol{\epsilon}_\lambda(\mathbf{q}_L^{(2)}), \boldsymbol{\epsilon}_\lambda(\mathbf{q}_L^{(3)})\}$. Following the same procedure for the mode $M_1^+$, we explored all the 15 non-symmetry equivalent configurations by independently distorting along the $\mathbf{Q}_L$ and $\mathbf{Q}_M$ subspaces. Combined distortions (mixing $\mathbf{Q}_L$ and $\mathbf{Q}_M$) were not considered. Each configuration was subsequently relaxed using DFT, with symmetry constraints imposed during relaxation.

In a second step, the relaxed structures were projected back onto the reduced order-parameter basis. This allowed for the classification of the resulting configurations beyond just the space-group symmetries. The projection is defined as
\begin{equation}
    Q_\lambda(\mathbf{q}^{(s)})=\sum_a\sum_i\sqrt{M_i}\,\boldsymbol{\epsilon}_{\lambda,i}^{*}(\mathbf{q}^{(s)})\cdot\Delta\mathbf{R}_{i,a}\cdot e^{-i\mathbf{q}^{(s)}\cdot\mathbf{T}_a}\,,
\end{equation}
which follows directly from Eq. \eqref{eq:dR} and the orthogonality of the polarization vectors. Configurations with symmetry-equivalent projections to $\mathbf{Q}_L$ and $\mathbf{Q}_M$, even if having non-identical energies, were grouped as the same local-minima. This is justified by the fact that such structural differences, of the order of picometers, are negligible when zero-point motion is included and tend to gain symmetries. This procedure reduces the Born-Oppenheimer energy landscape to six distinct local minima (not considering the pristine phase), including the well studied TrH and SoD reconstruction of the vanadium plane with different stacking sequences. Nevertheless, none of these minima is dynamically stable at a harmonic-level, reinforcing the necessity of beyond-harmonic approaches to describe the dynamics of CsV$_3$Sb$_5$. The original distorted geometries, relaxed structures, and their corresponding energies are listed in Table \ref{tab:Table2}.

The local-minima of $V(\mathbf{R})$ were studied including anharmonicity and ionic quantum fluctuations using the SSCHA \cite{monacelliStochasticSelfconsistentHarmonic2021}, this time without imposing symmetry constraints. The SSCHA relaxed structures were again projected back onto the order-parameter subspace $(\mathbf{Q}_L,\mathbf{Q}_M)$. Additionally, we considered a new configuration corresponding to a distortion along the mid-point of the LM phonon-branch. This structure, belonging to the P1 (No. 1) space-group, was computationally too demanding to be evaluated at a DFT level. In the case of this configuration, even thought the projection into the $LM$ phonon branch corresponds to a Cccm (No. 66) space-group, non negligible but small contributions of other phonons exist, preventing the system from acquiring point group symmetries. These calculations showed that zero point motion simplifies the free-energy landscape, $F[\rhos]$, into three metastable local-minima. In fact, just the TrH reconstruction of the kagome-layer was found dynamically stable, while the rest of the configurations (including the SoD) relaxed into it (see Table \ref{tab:Table2} and Table \ref{tab:Table3}). The presence of different local-minima is explained by stacking disorder between them. The structures which we define as TrH, $\pi$-TrH and 4TrH are illustrated in Fig. 1 in the main text, and their geometries provided as additional information.

\begin{table*}
\caption{\label{tab:Table2}Results on the possible independent charge reconstructions due to the dynamical instabilities $L_2^-$ and $M_1^+$ of the pristine phase. The order parameters describing the distortion are arranged such that the in-plane projections of the $\mathbf{q}_L^{(s)}$ and $\mathbf{q}_M^{(s)}$ wave vectors coincide. We note the arrangement of the Kagome layer into a TrH or SoD arrangement in each of the cases.}
\begin{ruledtabular}
\begin{tabular}{cc|cccc|cccc}
 \multicolumn{2}{c}{Initial structure}&\multicolumn{4}{c}{DFT relaxation}&\multicolumn{4}{c}{SSCHA relaxation at T=0 K}\\
 \makecell{Space\\ group}&$(\mathbf{Q}_L;\mathbf{Q}_M)$&\makecell{Space\\ group}&$(\mathbf{Q}_L;\mathbf{Q}_M)$&\makecell{$\Delta V$ \\ (meV/f.u.)}& V layer&\makecell{Space\\ group}&$(\mathbf{Q}_L;\mathbf{Q}_M)$&\makecell{$\Delta F$ \\ (meV/f.u.)}& V layer\\ \hline
 \multicolumn{8}{p{5cm}}{$1\times1\times1$}\\\hline
 No. 191 & (0, 0, 0; 0, 0, 0) & No. 191 & (0, 0, 0; 0, 0, 0) & 0 & Kagome & No. 191 & (0, 0, 0; 0, 0, 0)& 0& Kagome\\\hline
\multicolumn{8}{p{5cm}}{$2\times2\times1$}\\\hline
 No. 10 & (0, 0, 0; a, b, c) & No. 65 & (0, 0, 0; a, a, a) & -18.3 & TrH & No. 191 & (0, 0, 0; a, a, a)& -1.5& TrH\\
 No. 10 & (0, 0, 0; a, b, -c) & No. 65& (0, 0, 0; a, a, a) & -18.3 & TrH & No. 191 & (0, 0, 0; a, a, a)& -1.5& TrH\\
 No. 10 & (0, 0, 0; a, b, 0) & No. 65 & (0, 0, 0; a, a, a)& -18.3 & TrH & No. 191 & (0, 0, 0; a, a, a)& -1.5& TrH\\
 No. 47 & (0, 0, 0; a, 0, 0) & No. 47 & (0, 0, 0; a, 0, 0) & -4.9 & -- & No. 191 & (0, 0, 0; a, a, a)& -1.5 & TrH\\
 No. 65 & (0, 0, 0; a, a, 0) & No. 65 & (0, 0, 0; a, a, a) & -18.3 & TrH & No. 191 & (0, 0, 0; a, a, a)& -1.5&TrH\\
 No. 65 & (0, 0, 0; a, -a, 0) & No. 65 & (0, 0, 0; a, a, a) & -18.3 & TrH & No. 191 & (0, 0, 0; a, a, a)& -1.5& TrH\\
  No. 65 & (0, 0, 0; a, a, b) & No. 65 & (0, 0, 0; a, a, a) & -18.3 & TrH &No. 191 & (0, 0, 0; a, a, a)& -1.5& TrH\\
  No. 191 & (0, 0, 0; a, a, a) & No. 191 & (0, 0, 0; a, a, a) & -15.1 & TrH & No. 191 & (0, 0, 0; a, a, a)& -1.5& TrH\\
 No. 191 & (0, 0, 0; a, a, -a) & No. 191 & (0, 0, 0; a, a, -a) & -3.3 & SoD &No. 191 & (0, 0, 0; a, a, a)& -1.5& TrH\\\hline
 \multicolumn{8}{p{5cm}}{$2\times2\times2$}\\\hline
 No. 10 & (a, b, c; 0, 0, 0) & No. 69 & (a, a, 0; 0, 0, -a) & -25.2 & $\pi$-TrH & No. 69 & (a, a, 0; 0, 0, -a)& -2.3&$\pi$-TrH\\
 No. 12 & (a, b, 0; 0, 0, 0) & No. 69 & (a, a, 0; 0, 0, -a) & -25.2 & $\pi$-TrH&No. 69 & (a, a, 0; 0, 0, -a)& -2.3&$\pi$-TrH\\
 No. 65 & (a, a, b; 0, 0, 0) & No. 69 & (a, a, 0; 0, 0, -a)& -25.2 & $\pi$-TrH&No. 69 & (a, a, 0; 0, 0, -a)& -2.3&$\pi$-TrH\\
 No. 69 & (a, a, 0; 0, 0, 0) & No. 69 & (a, a, 0; 0, 0, -a) & -25.2 & $\pi$-TrH&No. 69 & (a, a, 0; 0, 0, -a)& -2.3&$\pi$-TrH\\
 No. 71 & (a, 0, 0; 0, 0, 0) & No. 71 & (a, 0, 0; 0, 0, 0) & -6.0 & -- &No. 69 & (a, a, 0; 0, 0, -a)& -2.3&$\pi$-TrH\\
 No. 191 & (a, a, a; 0, 0, 0) & No. 191 & (a, a, a; -b, b, -b) & -13.2 & SoD-TrH& No. 69 & (a, a, 0; 0, 0, -a)& -2.3&$\pi$-TrH\\

\end{tabular}
\end{ruledtabular}
\end{table*}

\begin{table*}
\caption{\label{tab:Table3} Results on an exceptional structure generated by a distortion along the midpoint of the unstable LM phonon branch. The order parameters describing the distortion are arranged such that the in-plane projections of the $\mathbf{q}_L^{(s)}$ and $\mathbf{q}_M^{(s)}$ wave vectors coincide. For the $\mathbf{q}_{LM}$ wave vector, the order parameters are organized into pairs of time-reversal-conjugate wave vectors, following the same in-plane arrangement as for $\mathbf{q}_L^{(s)}$ and $\mathbf{q}_M^{(s)}$, i.e., ($\mathbf{q}_{LM}^{(1)}, -\mathbf{q}_{LM}^{(1)}, \mathbf{q}_{LM}^{(2)}, -\mathbf{q}_{LM}^{(2)}, \mathbf{q}_{LM}^{(3)}, -\mathbf{q}_{LM}^{(3)}$).}
\begin{ruledtabular}
\begin{tabular}{cc|cccc}
 \multicolumn{2}{c}{Initial structure}&\multicolumn{4}{c}{SSCHA relaxation at T=0 K}\\
 Space group&$(\mathbf{Q}_L;\mathbf{Q}_M;\mathbf{Q}_{LM})$&Space group&$(\mathbf{Q}_L;\mathbf{Q}_M;\mathbf{Q}_{LM})$&$\Delta F$ (meV/f.u.)& V layer \\\hline
 \multicolumn{5}{p{5cm}}{$2\times2\times4$}\\\hline
 No. 1 & (0,0,0;0,0,0;a,a,a,0,0,0) & No. 1 & (b,-b,-2b;-b,b,0;b,b,b,b,0,0) & -2.2 & 4TrH\\
 \end{tabular}
\end{ruledtabular}
\end{table*}

\section{Anharmonic Spectral Function}

Within the dynamical formulation of the SSCHA theory~\cite{BiancoSecondOrderStructural2017}, the phonon spectral function is derived from the retarded one-phonon Green function, defined as
\begin{equation}
\mathbf{G}(\mathbf{q}, \omega) = \left[ (\omega + i\eta)^2 \mathbf{I} - \boldsymbol{\omega}^2(\mathbf{q}) - \boldsymbol{\Pi}(\mathbf{q}, \omega + i\eta) \right]^{-1}.
\end{equation}
Above, $\mathbf{q}$ is the phonon wave vector, $\boldsymbol\omega(\mathbf{q})$ the SSCHA auxiliary dynamical matrix in the mode basis and $\eta$ a positive infinitesimal. $\boldsymbol{\Pi}(\mathbf{q}, \omega)$ is the phonon-phonon self-energy matrix, which is non-zero in the presence of anharmonicity.

In the current implementation, the self-energy is approximated within the bubble approximation into 
\begin{equation}
    \mathbf{\Pi}(\mathbf{q},\omega)\approx\mathbf{\Pi}^{(B)}(\mathbf q, \omega)=\mathbf{\Phi}^{(3)}(\mathbf{q}):\mathbf{\Lambda}(\omega):\mathbf{\Phi}^{(3)}(\mathbf{q})\,,
\end{equation}
where $\mathbf{\Phi}^{(3)}(\mathbf{q})$ are the third order force constants, and $\mathbf{\Lambda}(\omega)$ is a fourth order tensor dependent on the SSCHA auxiliary frequencies and polarization vectors. For simplicity we used the compacted formalism, where $\mathbf{X}:\mathbf{Y}$ stands for the double summation in the last two indices of $\mathbf{X}$ and the first two indices of $\mathbf{Y}$.

In order to overcome convergence issues in the calculation of $\mathbf{\Pi}^{(B)}(\mathbf{q},\omega)$ a small but finite smearing $\eta$ is employed along with a interpolated fine phonon $\mathbf{k}$-grid. This is mathematically expressed as

\begin{equation}
\begin{split}
    \mathbf{\Pi}^{(B)}(\mathbf{q},\omega+ i\eta)=\sum_{\mathbf{k}_1\mathbf{k}_2}\sum_\mathbf{G}\delta_{\mathbf{G},\mathbf{q}+\mathbf{k}_1+\mathbf{k}_2}\cdot\\
    \cdot\mathbf{\Phi}^{(3)}(-\mathbf{q},-\mathbf{k}_1,-\mathbf{k}_2):\mathbf{\Lambda}(\omega+ i\eta,-\mathbf{k}_1,-\mathbf{k}_2,\mathbf{k}_1,\mathbf{k}_2):\\ \mathbf{\Phi}^{(3)}(\mathbf{q},\mathbf{k}_1,\mathbf{k}_2)\,,
\end{split}
\end{equation}
where $\mathbf{k}_i$ are phonon wave vectors, and $\mathbf{G}$ is a reciprocal lattice vector. 

In the general case, the self-energy $\boldsymbol{\Pi}^{(B)}(\mathbf{q}, \omega)$ is a Hermitian matrix with components $\Pi^{(B)}_{\mu\nu}(\mathbf{q}, \omega)$ in the phonon mode basis, where $\mu$ and $\nu$ label the vibrational branches at wave vector $\mathbf{q}$. Therefore, in the presence of anharmonicity, hybridization occurs, and vibrational modes are no longer well defined quasiparticles. This is evident when writing the spectral function
\begin{equation}
\mathbf{A}(\mathbf{q}, \omega) = -\frac{\omega}{\pi} \operatorname{Im}  \left[ \operatorname{Tr} \mathbf{G}(\mathbf{q}, \omega) \right] ,
\end{equation}
which contains all information on the energies and lifetimes of phonon excitations, without any mode index labeling.

In order to combine the SSCHA dynamical theory with density functional perturbation theory electron-phonon calculations, we need to keep track of the contribution to the spectra of each vibrational mode. This is achieved under the no mode-mixing approximation. By neglecting the off-diagonal elements $\Pi^{(B)}_{\mu\nu}(\mathbf{q}, \omega)$ with $\mu \ne \nu$, and assuming that each phonon mode propagates independently, the Green function then becomes diagonal in the mode index $\mu$. Consequently, the mode-resolved spectral function simplifies to
\begin{equation}
A_{\mu}(\mathbf{q}, \omega) = -\frac{1}{\pi} \operatorname{Im} \left[ \frac{1}{(\omega + i\eta)^2 - \omega_{\mu}^2(\mathbf{q}) - \Pi_{\mu\mu}(\mathbf{q}, \omega)} \right]\,,
\end{equation}
which can be expressed with good approximation as a Lorentzian function
\begin{equation}
    A_\mu(\mathbf{q},\omega)\approx\frac{1}{\pi}\frac{\Gamma_\mu(\mathbf{q})}{[\omega-\Omega_\mu(\mathbf{q})]^2+[\Gamma_\mu(\mathbf{q})]^2}\,,
\end{equation}
centered at the frequency
\begin{equation}
    \Omega_\mu(\mathbf{q})=\mathrm{Re}\left\{\sqrt{\omega_\mu^2(\mathbf{q})+\Pi_{\mu\mu}(\mathbf{q},\omega})\right\}\,,
\end{equation}
with a half width at half maximum (HWHM) of
\begin{equation}
    \Gamma_\mu(\mathbf{q})=-\mathrm{Im}\left\{\sqrt{\omega_\mu^2(\mathbf{q})+\Pi_{\mu\mu}(\mathbf{q},\omega)}\right\}\,.
\end{equation}
This approximation is useful when combining anharmonic spectral functions with DFPT electron-phonon linewidths. Knowing the HWHM $\Gamma_\mu^{\mathrm{el-ph}}(\mathbf{q})$ due to electron-phonon interaction, the total linewidth of mode $\mu$ is approximated to a sum of independent contributions of electron-phonon and phonon-phonon scattering events, i.e.
\begin{equation}
    \Gamma_\mu(\mathbf{q})=\Gamma_\mu^{ph-ph}(\mathbf{q})+\Gamma_\mu^{el-ph}(\mathbf{q})\,.
\end{equation}
The phonon spectra employed in the calculation of the anharmonic spectral function was computed in a $2\times2\times2$ $\mathbf{q}$-grid of the $\pi$-TrH phase, where in the calculation of the $\mathbf{\Pi}^{(B)}(\mathbf{q},\omega)$, the phonon-spectra was interpolated into a $4\times4\times4$ fine $\mathbf{k}$-grid with a smearing value of $\eta$=3 cm$^{-1}$. The computational approach of the DFPT electron-phonon calculations is described below.
\section{Softening of the $T_1^+$ phonon-mode under pressure}
\begin{figure}
\includegraphics[width=0.5\textwidth]{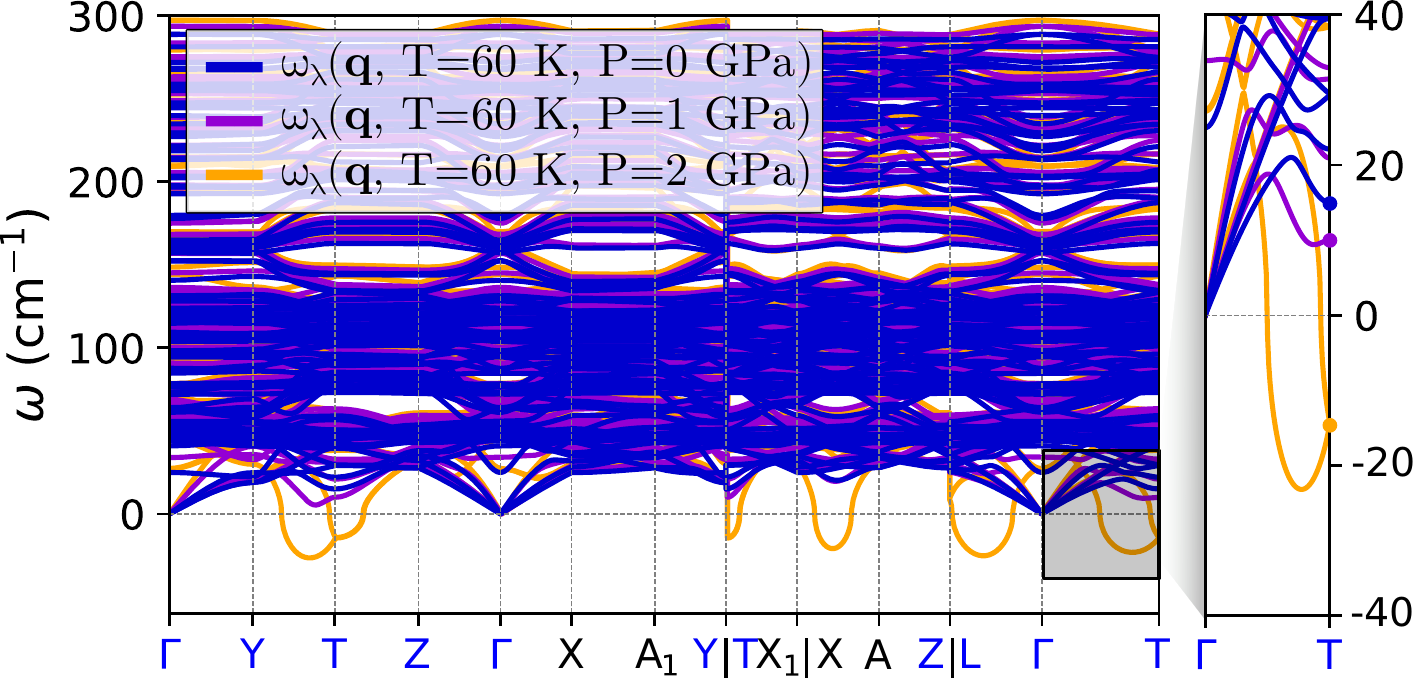}% Here is how to import EPS art
\caption{\label{fig:S2} Phonon-spectra of the $\pi$-TrH phase at T=60 K under pressure. In the right panel, a zoom in of the $\Gamma$T path shows the phonon softening of the $T_1^+$ phonon-mode under pressure. The high symmetry wave vectors calculated explicitly and  later used for the Fourier interpolation are highlighted in blue.}
\end{figure}

As discussed in the main text, CsV$_3$Sb$_5$ exhibits a lattice-instability transforming under the irreducible representation of $T_1^+$ at approximately 80 K. The $T_1^+$ mode is non-degenerate, and the resulting distorted structure belongs to the Cmmm (No. 65) space-group with a $4\times4\times2$ modulation of the high-symmetry phase. Although this instability might suggest the existence of an intermediate phase between 80 and 89 K, such a phase has not being observed experimentally \cite{frachetColossalAxisResponse2024}. However, it could be related to a second CDW phase with a larger in-plane order that has been recently observed under compression \cite{zhengEmergentChargeOrder2022, stierPressureDependentElectronicSuperlattice2024}. To corroborate such an hypothesis, the renormalization of the $T_1^+$ phonon-mode under pressure was calculated at T=60 K. The results show that this mode softens with pressure and collapses around 1.5 GPa (Fig. \ref{fig:S2}). This finding links the in-plane super lattice observations under isotropic pressure with the collapse of the $T_1^+$ phonon mode \cite{zhengEmergentChargeOrder2022, stierPressureDependentElectronicSuperlattice2024}, even suggesting that the in-plane super lattice could play a role in the CDW transition.

\section{Stacking disorder induced splitting in phonon and electronic bands}
\begin{figure}
\includegraphics[width=0.4\textwidth]{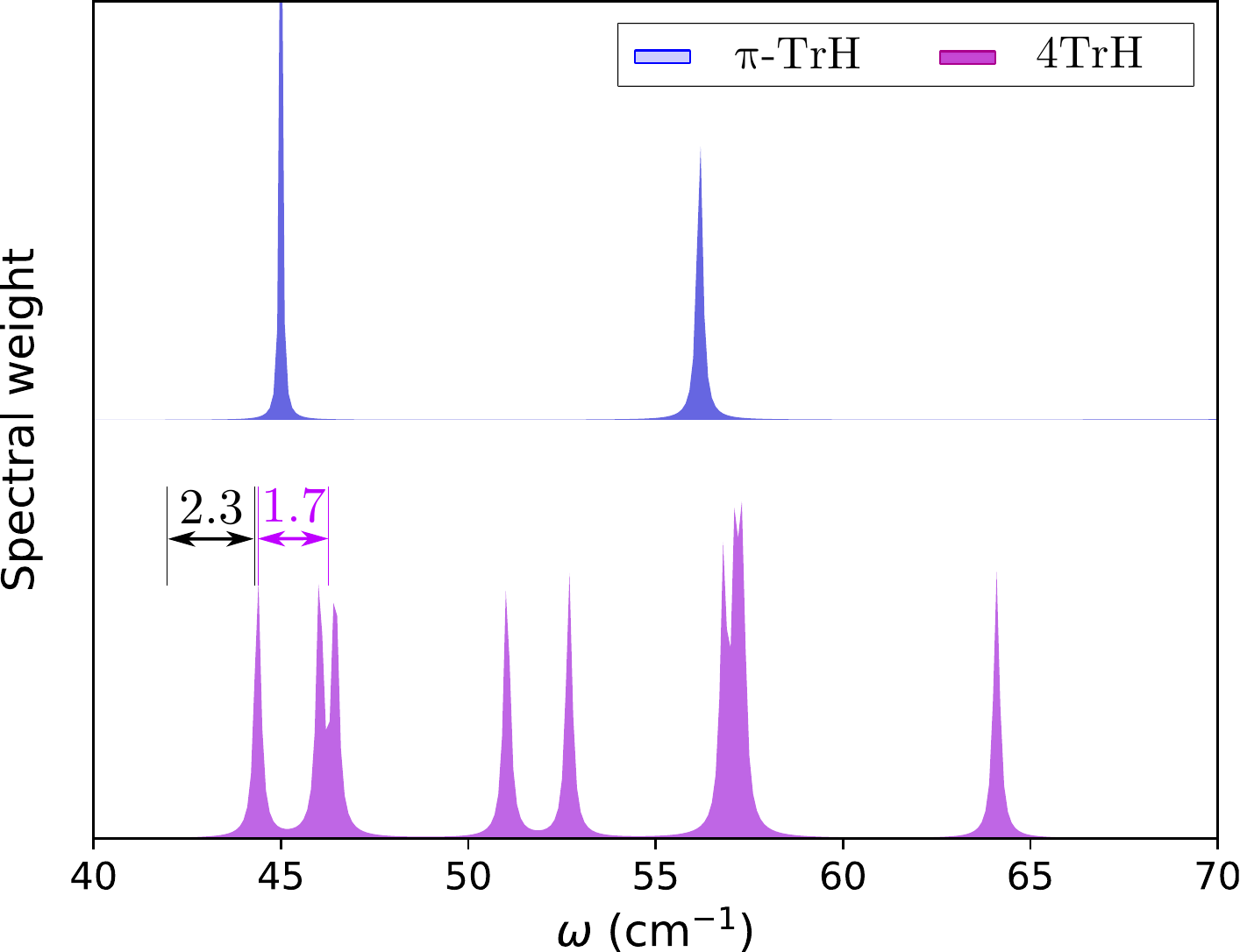}% Here is how to import EPS art
\caption{\label{fig:S4} Phononic spectral function of the $\pi$-TrH and 4TrH phase considering $\Gamma_1^+$ modes. The experimental (black) and calculated (violet) splittings are highlighted. For the $\pi$-TrH phase, and within the no mode-mixing approximation, the spectral peaks are approximated into Lorentzians including phonon-phonon and electron-phonon interactions. For the 4TrH phase, a reasonable constant value of 0.1 cm$^{-1}$ is cosidered for the HWHM of the Lorentzians, and the Cccm (No. 66) space-group was considered for the filtering of the $\Gamma_1^+$ modes. The TrH phase is not considered as it does not have any $\Gamma_1^+$ spectral weight in this energy window. For visualization, an offset is included to the $\pi$-TrH spectra.}
\end{figure}

Symmetry breaking induced by stacking disorder naturally leads to the splitting of both phonon modes and electronic bands. This phenomena has been observed experimentally, raising doubts about the adequacy of the TrH vanadium reconstruction alone for a complete description of the CDW phase. Recent ultrafast time-resolved reflectivity measurements have revealed a splitting in the lowest energy $\Gamma_1^+$ phononic spectral peak of CsV$_3$Sb$_5$ in the CDW phase \cite{liangCoherentPhonon2025}, which cannot be explained by the $\pi$-TrH phase alone. In Fig.~\ref{fig:S4}, we compare our calculated anharmonic phonon spectra for the CDW phase with these experimental results. We show that in a stacking-disordered structure, such as the 4TrH phase, a splitting of 1.7 cm$^{-1}$ emerges, fully consistent with the experiments, and thus supporting stacking disorder as the underlying cause of the observed splitting.

\begin{figure*}
\includegraphics[width=\textwidth]{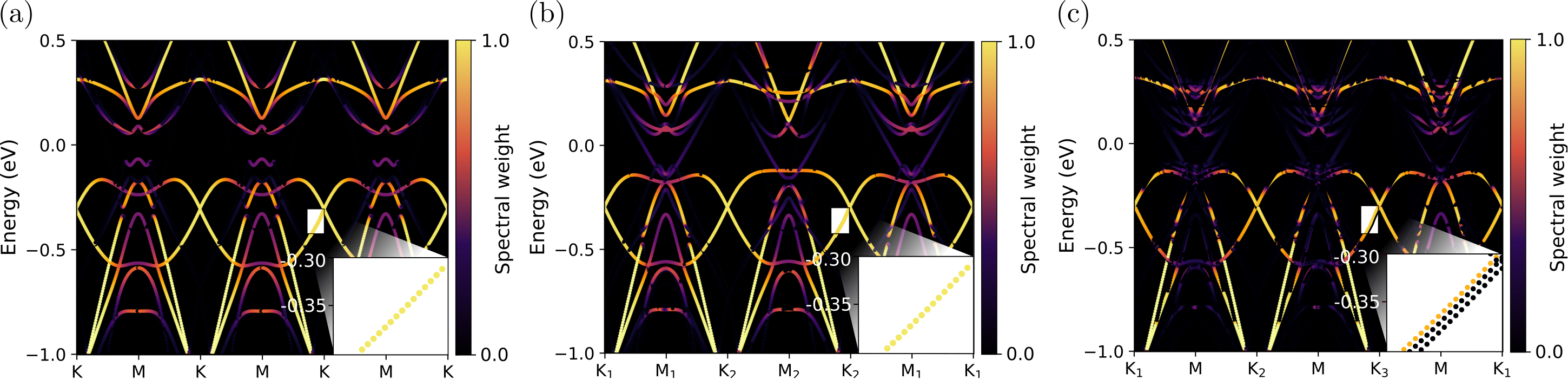}% Here is how to import EPS art
\caption{\label{fig:S3} (a) Unfolded electronic band structure of the TrH phase along the in-plane high-symmetry path represented Fig. 3 of the main text by arrows. The $\mathbf{k}$-path is given respect to the high-symmetry Brillouin zone. The inset corresponds to a zoom-in in order to visualize if band-splitting is present in the KM direction. (b) and (c) same as (a) for the $\pi$-TrH and the 4TrH phases respectively.}
\end{figure*}

Similar phenomena has been reported for in-plane electron bands in ARPES measurements \cite{kangChargeOrderLandscape2023}. Therefore, along with the fermiology, we calculated the in-plane band dispersion for the distinct stacking phases, unfolding the bands to the Brillouin-zone of the high-symmetry phase to facilitate direct comparison with experiments. The results, shown in Fig. \ref{fig:S3}, highlight the region where band splitting has been experimentally observed. In agreement with experimental findings, the 4TrH phase reproduces the band splitting along the KM direction, whereas it is absent for other phases. Therefore, not only in-plane transport isotropy, but also band renormalization can be understood considering stacking disorder of vanadium TrH layers. The splitting is however smaller than the one observed experimentally.

\section{Combining the DFPT deformation potential with anharmonic phonons}

The electron-phonon matrix elements employed for the estimation of the electron-phonon linewidth, as well as for superconductivity, were computed using density functional perturbation theory (DFPT) implemented in the Quantum Espresso package \cite{Giannozzi_2009,Giannozzi_2017}. We used energy cutoffs of 80/800 Ry for the wavefunctions/density together with a Methfessel-Paxton smearing of 0.01 Ry. A $6\times6\times6$ and a $16\times16\times8$ $\mathbf{k}$-grid were used for the sampling of the Brillouin zone in the $\pi$-TrH and the high symmetry phase respectively. The employed ultra soft pseudopotentials were generated by Dal Corso \cite{DALCORSO2014337}, with valence configurations of 5s$^2$5p$^6$6s$^1$, 3s$^2$3p$^6$3d$^3$4s$^2$, and 5s$^2$5p$^3$ for Cs, V and Sb respectively. We used the optB88-vdW exchange-correlation functional \cite{klimesChemicalAccuracyVan2009}, previously proved to accurately capture the Van der Waals character of the material.

Within this framework, the electron-phonon matrix elements are given by
\begin{equation}
\begin{split}
    g^{\lambda}_{n,m}(\mathbf{k},\mathbf{k}+\mathbf{q})=\sum_{\alpha,i}\frac{1}{\sqrt{2M_i\omega_\lambda(\mathbf{q}})}\epsilon_{\lambda,i}^{\alpha}\\\langle n\mathbf{k}| \left[ \frac{\partial V_{KS}}{\partial u_i^\alpha(\mathbf{q})} \right]_0|m\mathbf{k}+\mathbf{q}\rangle\,,
\end{split}
\end{equation}
where $\lambda$ is the vibrational mode index, $n$ and $m$ label the electronic states, and $\mathbf{k}$ and $\mathbf{q}$ are the wave vectors of the electronic state and phonon respectively. $M_i$ is the mass of the atom with atomic index $i$, $u_i^\alpha(\mathbf{q})$ is a displacement of atom $i$ along the cartesian direction $\alpha$. $\epsilon_{\lambda,i}^\alpha$ is the polarization vector of the mode $\lambda$, while $\partial V_{KS}/\partial u_i^\alpha(\mathbf{q})$ is the so-called deformation potential.

Considering that all relevant physics occur at the Fermi energy, the Eliashberg function can be averaged at the Fermi surface. Moreover, considering T=0 K (no excited vibrational modes), the Eliashberg function is defined as
\begin{equation}
\begin{split}
        \alpha^2F(\omega)=\frac{1}{N(\varepsilon_F)}\frac{1}{N_\mathbf{k}}\frac{1}{N_\mathbf{q}}\sum_{\mathbf{k}\mathbf{q}}\sum_\lambda\sum_{nm} \left|g^\lambda_{n\mathbf{k}, m\mathbf{k}+\mathbf{q}}\right|^2 \\
        \delta(\omega-\omega_\lambda(\mathbf{q}))\delta(\varepsilon_{n\mathbf{k}}-\varepsilon_F)\delta(\varepsilon_{m\mathbf{k}+\mathbf{q}}-\varepsilon_F)\,.
\end{split}
\end{equation}
$N(\varepsilon_F)$ is the density of states at the Fermi level per spin, $\varepsilon_F$ is the Fermi energy, and $N_\mathbf{k}$ and $N_\mathbf{q}$ are the number of wave vectors employed in the sampling of the Brillouin zone for electrons and phonons respectively. By defining 
\begin{equation}
\begin{split}
    \Delta^{\alpha\beta}_{ab}(\mathbf{q})=\frac{1}{N(\varepsilon_F)}\frac{1}{N_\mathbf{k}}\sum_\mathbf{k}\sum_{nm}\langle n\mathbf{k}|\frac{\partial V_{KS}}{\partial u^\alpha_a(\mathbf{q})}|m\mathbf{k}+\mathbf{q}\rangle\\
    \langle m\mathbf{k}+\mathbf{q}|\frac{\partial V_{KS}}{\partial u^\beta_b(\mathbf{q})}|n\mathbf{k}\rangle
    \delta(\varepsilon_{n\mathbf{k}}-\varepsilon_F)\,\delta(\varepsilon_{m\mathbf{k}+\mathbf{q}}-\varepsilon_F)\,,
\end{split}
\end{equation}
the Eliashberg function can be rewritten as
\begin{equation}
    \begin{split}
                \alpha^2F(\omega)=\frac{1}{N_\mathbf{q}}\sum_{\mathbf{q}\lambda}\sum_{ab}\sum_{\alpha\beta}\frac{\varepsilon^\alpha_{\lambda,a}(\mathbf{q})\Delta^{\alpha\beta}_{ab}(\mathbf{q})\varepsilon_{\lambda,b}^\beta(-\mathbf{q})}{2\omega_\lambda\sqrt{M_aM_b}}\\
                \delta(\omega-\omega_\lambda(\mathbf{q}))\,.
    \end{split}
\end{equation}

Note that $\Delta^{\alpha\beta}_{ab}(\mathbf{q})$ contains all information regarding electron-phonon coupling, while it is independent of the phonon-spectra of the system. This allows the description of anharmonic systems, by combining the DFPT calculated $\Delta^{\alpha\beta}_{ab}(\mathbf{q})$ with phonon-spectra obtained within the SSCHA framework. The phonon spectra employed in the estimation of $T_c$ were the SSCHA auxiliary phonon modes calculated in a $2\times2\times2$ $\mathbf{q}$-grid for both the $\pi$-TrH and high symmetry phases. In the low symmetry phase, during the non self-consistent calculation of $\Delta_{ab}^{\alpha\beta}(\mathbf{q})$, a $8\times8\times8$ $\mathbf{k}$-grid was used together with a Gaussian smearing of 0.01 Ry. In the pristine phase, the non self-consistent calculation was computed in a equivalent $\mathbf{k}$-grid $16\times16\times8$, while the deformation potential  as well as the auxiliary phonon-spectra $\omega_\lambda(\mathbf{q})$ were interpolated into a $4\times4\times2$ $\mathbf{q}$-grid.

The superconducting critical temperature was estimated with the Allen-Dynes modified McMillan equation with the use of a Coulomb pseudopotential of $\mu^*=0.15$. Values obtained with other values of $\mu^*$ are given in Table \ref{tab:Table4}.
\begin{table*}
\begin{caption}
    {\label{tab:Table4}Results on the density functional perturbation theory calculated superconducting transition of the $\pi$-TrH and high-symmetry phases. For the $\pi$-TrH phase, the phonon spectra was calculated at T=0 K, while for the pristine phase 100 K were required to stabilize the spectra. In the calculation of $T_c$, McMillan Allen-Dynes modified equation was used.}
\end{caption}
\begin{ruledtabular}
\begin{tabular}{c | c c c c c}
Phase & $\lambda$ & $\omega_{\log}$  & $T_c$ ($\mu^*=0.10$) & $T_c$ ($\mu^*=0.15$) & $T_c$ ($\mu^*=0.20$) \\
\hline
$\pi$-TrH & 0.68 & 88.48 (cm$^{-1}$) & 4.4 K & 2.7 K& 1.4 K \\
High symmetry & 2.03 & 54.18 (cm$^{-1}$) & 14.1 K & 12.1 K& 10.2 K \\
\end{tabular}
\end{ruledtabular}
\end{table*}
\bibliography{CsV3Sb5_paper}